\documentclass[preprint]{sigplanconf}

\pdfoutput=1
\preprintfooter{PEPM~'14, January 20--21, 2014, San Diego, CA, USA.}

\usepackage{booktabs}
\usepackage{threeparttable}
\usepackage{microtype}
\usepackage[hidelinks]{hyperref}
\usepackage{graphicx}
\usepackage[utf8]{inputenc}

\newcommand{\lineref}[1]{(\texttt{#1})}
\newcommand{\ub}[0]{\_\allowbreak}

\begin{document}

\setlength{\pdfpageheight}{\paperheight}
\setlength{\pdfpagewidth}{\paperwidth}

\exclusivelicense
\conferenceinfo{PEPM~'14}{January 20--21, 2014, San Diego, CA, USA}
\copyrightyear{2014} 
\copyrightdata{978-1-4503-2619-3/14/01}
\doi{2543728.2543733}

\title{QEMU/CPC:\\Static Analysis and CPS Conversion\\for Safe,
  Portable, and Efficient Coroutines}

\authorinfo{Gabriel Kerneis\and Charlie Shepherd}
{University of Cambridge}
{\{gk338,cs648\}@cam.ac.uk}
\authorinfo{Stefan Hajnoczi}
{Red Hat}
{stefanha@redhat.com}

\maketitle

\begin{abstract}
Coroutines and events are two common abstractions for writing concurrent
programs.  Because coroutines are often more convenient, but events more
portable and efficient, it is natural to want to translate the former
into the latter.  CPC is such a source-to-source translator for C
programs, based on a partial conversion into continuation-passing style
(CPS conversion) of functions annotated as cooperative.

In this article, we study the application of the CPC translator to QEMU,
an open-source machine emulator which also uses annotated coroutine
functions for concurrency.  We first propose a new type of annotations
to identify functions which never cooperate, and we introduce CoroCheck, a tool for
the static analysis and inference of cooperation
annotations.  Then, we improve the CPC translator, defining CPS
conversion as a calling convention for the C language, with support for
indirect calls to CPS-converted function through function pointers.
Finally, we apply CoroCheck and CPC to QEMU (750\,000 lines of C code),
fixing hundreds of missing annotations and comparing performance of the
translated code with existing implementations of coroutines in QEMU.

Our work shows the importance of static annotation checking to prevent
actual concurrency bugs, and demonstrates that CPS conversion is a
flexible, portable, and efficient compilation technique, even for very
large programs written in an imperative language.
\end{abstract}




\section{Introduction}
\label{sec:intro}

Most computer programs are \emph{concurrent} programs, which need to perform
several tasks at the same time.  For example, a network server needs to serve
multiple clients at a time; a GUI needs to handle multiple keyboard and mouse
inputs; and a network program with a graphical interface (e.g.\ a
virtual machine with an emulated network card) needs to do both
simultaneously.

\paragraph{Threads and events}

There are many different techniques to implement concurrent programs.  A
very common abstraction is provided by \emph{threads}.
In a threaded program, concurrent tasks are executed
by a number of independent threads which communicate through a shared
memory heap.  Threads are generally either \emph{native threads},
preemptively scheduled by the Operating System (OS), or \emph{user-space
threads}, cooperatively scheduled by a library.

An alternative to threads is \emph{event-driven} programming.
An event-driven program interacts with its environment by reacting to a
set of stimuli called \emph{events}.  At any given point in time, to
every event is associated a piece of code known as the \emph{handler}
for this event.  A global scheduler, known as the \emph{event loop},
repeatedly waits for an event to occur and invokes the associated
handler.  Performing a complex task requires to coordinate several event
handlers by exchanging appropriate events.

Unlike threads, event handlers do not have an associated stack;
event-driven programs are therefore more lightweight and often faster
than their threaded
counterparts~\cite{ousterhout1996threads,DBLP:conf/sosp/WelshCB01}.
They are also more portable than native threads, because they do not
require OS support for task switching.  However, because it splits the
flow of control into multiple tiny event handlers, event-driven
programming is generally deemed more difficult and
error-prone~\cite{DBLP:conf/usenix/AdyaHTBD02}, in particular in
imperative languages such as C with no support for closures and
first-class functions.  Additionally, because of its cooperative nature,
event-driven programming alone is often not powerful enough, in
particular when accessing blocking APIs or using multiple processor
cores; it is then necessary to write \emph{hybrid} code, that uses both
native threads and event handlers, which is even more difficult.

\paragraph{Continuation-Passing C}

Since event-driven programming is more difficult but more efficient than
threaded programming, it is natural to want to at least partially
automate it.  \emph{Continuation-Passing C} (CPC~\cite{hosc2012}) is an
extension of the C programming language for writing concurrent systems,
built on top of the C Intermediate Language (CIL) framework~\cite{DBLP:conf/cc/NeculaMRW02}.
The CPC programmer manipulates very lightweight threads, annotating
cooperative functions and choosing whether they should be cooperatively
or preemptively scheduled at any given point.  The CPC program is then
compiled in two steps: it is first
processed by the \emph{CPC translator}, which produces highly-efficient
sequentialized event-driven code, and then linked with the \emph{CPC
runtime}, a small optimised C library scheduling
the continuations introduced by the CPC translator.
The translation from annotated cooperative functions
into events is performed by a series of classical source-to-source
program transformations: splitting of the control flow into mutually
recursive nested functions, lambda lifting of these functions,
and CPS conversion.  This
approach retains the best of both worlds: the relative convenience of
programming with threads, and the low memory usage of event-loop code.

\paragraph{The QEMU emulator}

\emph{Quick EMUlator} (QEMU~\cite{DBLP:conf/usenix/Bellard05}) is an
open-source machine emulator and virtualizer that supports 16~CPU
architectures and many individual devices including network cards,
storage controllers, and graphics cards.  Code execution is done either
through dynamic translation, or through hardware virtualization support
available in modern x86 CPUs.  It is a large and complex project:
750\,000 lines of code written by 645~contributors over more than
10~years.

Running a guest system with QEMU involves executing guest code, handling
timers, processing I/O, and responding to the management console. Performing
all these tasks at once requires an architecture capable of mediating
resources in a safe way without pausing guest execution if a disk I/O or
a command from the management console takes a long time to complete.
QEMU uses a hybrid architecture that combines event-driven programming
with threads.

To simplify the event-driven part, QEMU uses \emph{coroutines}.
Coroutines are an old control abstraction, commonly characterised by the
ability to resume and suspend execution, as well to preserve the values of
local data between successive calls.  Among the many existing styles of
coroutines~\cite{DBLP:journals/toplas/MouraI09}, QEMU implements first-class,
stackful, asymmetric coroutines, with no value passing upon suspension and
resumption.  As a result, programming with QEMU coroutines feels a lot
like programming with threads, with the exception that cooperating
passes control back to the parent coroutine instead of some global
scheduler.  Each coroutine is responsible for registering itself with
the main event loop before yielding.

Similarly to CPC, QEMU functions that should be executed within
coroutines are annotated.  These annotations are not intended to drive a
source-to-source transformation with static checking, but they are an
essential piece of documentation for developers to write correct coroutine
code.  However, they are never statically checked.

Coroutines in QEMU are implemented in a platform-dependent manner.
There are currently two classes of coroutine implementations:
\emph{stack-switching backends}, which allocate a new stack for each
    coroutine and perform a context switch when entering and
    yielding; and
\emph{thread-based backends}, which create a new thread for each
    coroutine and use synchronization primitives to ensure that only one
    coroutine runs at a given time and that control is transferred in
    the correct order.
The former use non-portable functions, such as \verb+sigaltstack+ and
\verb+swapcontext+ on Unix, and \verb+SwitchToFiber+ on Windows: the
context switch is triggered from user-space, and in most cases involves
only a cheap function call to set the current stack pointer.  The latter
is built on top of GLib's threads, a more portable but slower approach.

These approaches can be made to work well but require maintenance and
a relatively high porting effort when targeting new platforms.
Consequently, new QEMU ports sometimes only use the slower
thread-based backends.  The stack-switching backends require in-depth
knowledge of the CPU architecture and low-level support for switching
the runtime stack and process context such as signal masks.

The maintenance cost of coroutine backends, and the lack of static
verification of coroutine annotations, create a need for a new approach
that offers safety guarantees, and good portability with no performance
degradation.

Our approach in this paper is to process the whole QEMU source code with
the CPC translator in order to convert it to continuation-passing style,
and then link the resulting code with a new, portable coroutine backend
for QEMU based on continuations.  Therefore, we reuse the CPC
translator, but not the original CPC runtime: the point of our work is
indeed not to rewrite QEMU in the CPC language, but to use the CPC
compilation technique to implement the coroutine API defined by QEMU.

\subsection*{Overview of the paper}

This article is a case study in applying CPS conversion and static
analysis on a large C program, to implement safe, portable and efficient
coroutines.

\paragraph{Contributions}
We make the following contributions:
\begin{enumerate}
  \item a new type of annotations to mark blocking functions;
  \item CoroCheck, a tool for the static analysis and inference of
    cooperation and blocking annotations, used to rectify hundreds of
    annotations in QEMU, a real-world open-source project of
    over 750\,000 lines of C code;
  \item a performance comparison of continuation-based coroutines to
    several other existing implementations of coroutines for QEMU.
\end{enumerate}
Our work shows the importance of static annotation checking to prevent
actual concurrency bugs, and demonstrates that CPS conversion --- as
implemented by the CPC translator --- is a flexible, portable, and
efficient compilation technique, even for very large programs written in
an imperative language.

\paragraph{Outline}

We first give an overview of related work
(Section~\ref{sec:related-work}), and QEMU coroutines
(Section~\ref{sec:qemu-api}).  Then, we introduce CoroCheck, our tool
for the static analysis and inference of coroutine annotations
(Section~\ref{sec:corocheck}).  Next, we present the CPC transformation
technique (Section~\ref{sec:cpc}), and dive into the challenges
associated with applying CPS conversion to QEMU
(Section~\ref{sec:cpc-qemu}).  Finally, we evaluate performance results
(Section~\ref{sec:eval}), and conclude (Section~\ref{sec:conclusion}).

\paragraph{Timeline}
The work described has been carried out over the course of three months,
in the context of a Google Summer of Code project.  The second author, who had
no prior knowledge of either QEMU or CPC, is the sponsored student. The first
and third authors co-mentored his work, respectively for the CPC and QEMU sides
of the project.  The timeline went roughly as follows:
\begin{description}
  \item[Before the begining of the project] The first author spent two weeks
    improving CIL (Section~\ref{sec:cil-bugs}) and CPC
    (Section~\ref{sec:cps-calling-convention}).
  \item[First month] The second author studied the implementation of the CPC
    translator and runtime, QEMU coroutines, and implemented a prototype of the
    \texttt{coroutine-cpc} backend (Section~\ref{sec:cpc-backend}).
  \item[Second month] The second author started adding missing coroutine
    annotations in the block layer (Section~\ref{sec:missing-annot}), to
    compile two small, stand-alone utilities provided by QEMU
    (\texttt{qemu-img} and \texttt{qemu-io}). To support his refactoring
    effort, the first author spent a few days writing a first prototype of
    CoroCheck, a tool for the static analysis of coroutine
    annotations (Section~\ref{sec:coro-features}), and two more weeks
    adding a plug-in mechanism to CIL and rewriting CoroCheck as a CIL plug-in.
  \item[Third month] The second author published two series of patches adding
    coroutine annotations to QEMU, and improved them with the help of the third
    author and other QEMU developers.  After some minor clean-up and
    optimisations, the first author was able to run a virtual machine using
    QEMU with a \texttt{coroutine-cpc} backend, and perform micro- and
    macro-benchmarks (Section~\ref{sec:eval}).
\end{description}

\paragraph{Software availability}

The code developed as part of this work is available as free
software.  Whenever possible, the changes have been integrated directly
in the original projects that we had to adapt (CIL~\cite{cil-git},
CPC~\cite{cpc-git}, QEMU~\cite{qemu-git}).  Our new tool CoroCheck is
also available in its own repository~\cite{coro-git}.  Some patches for
the CPC backend of QEMU\footnote{Available online at
\url{http://github.com/kerneis/qemu/}.} are still pending review by the
QEMU team at the time of writing.

\section{Related work}
\label{sec:related-work}

\paragraph{Continuations and concurrency}

Delimited continuations are the general abstraction to think of threads,
events and coroutines. A delimited continuation can be realized in many
ways: as a stack implicitly associated with a thread, as an explicitly
copied part of the stack, as a sequence of activation frames stored on
heap or in a data structure (as happens in continuation-passing style),
or as a closure (an event handler).  In functional languages,
thread-like primitives are commonly built either on top of first-class
(delimited) continuations, or encapsulated within a continuation monad.

The former approach is best illustrated by Concurrent ML
constructs~\cite{DBLP:conf/mcmaster/Reppy93}, implemented on top of SML/NJ's
first-class continuations, or by the way coroutines are typically implemented
in Scheme using the \texttt{call/cc}
operator~\cite{DBLP:journals/cl/HaynesFW86}. More recently, Scala uses
first-class delimited continuations to implement concurrency
primitives~\cite{DBLP:conf/icfp/RompfMO09,DBLP:journals/tcs/HallerO09}.
Anton and Thiemann build pure OCaml
coroutines~\cite{DBLP:conf/aplas/AntonT10} on top of Kiselyov's delimcc
library for delimited continuations~\cite{DBLP:journals/tcs/Kiselyov12}.

Explicit translation into continuation-passing style, often encapsulated
within a monad, is used in languages lacking first-class continuations.
In Haskell, Claessen proposes a monad transformer yielding a concurrent
version of existing monads~\cite{DBLP:journals/jfp/Claessen99}. Li and
Zdancewic also use a monadic approach to build event-driven network
servers~\cite{DBLP:conf/pldi/LiZ07}.   In OCaml, Vouillon's
Lwt~\cite{DBLP:conf/ml/Vouillon08} provides a lightweight alternative to
native threads.  The asynchronous model in F\# is implemented with a
localized continuation-passing translation of control-flow and a
heap-based allocation of the closures, using three continuations for
success, exceptions and cancellation~\cite{DBLP:conf/padl/SymePL11}.

\paragraph{From threaded to event-driven style}

In imperative languages, first-class continuations are generally not
available and monadic style extremely inconvenient.  This makes program
transformation techniques more widespread, with two main approaches:
translating loops and gotos into state machines~\cite{bohm}, or
converting functions into continuation-passing
style~\cite{strachey,DBLP:journals/tcs/Plotkin75}.

Deriving state-machines from a threaded-style code is as old as
\emph{Duff's device}~\cite{duff}.  Implementations have then been
improved in multiple directions: as C preprocessor
macros~\cite{DBLP:conf/sensys/DunkelsSVA06}, as source-to-source
transformations on C++~\cite{DBLP:conf/usenix/KrohnKK07} or
Java~\cite{DBLP:conf/pepm/FischerMM07} programs, as a transformation on
JVM bytecode~\cite{DBLP:conf/ecoop/SrinivasanM08}, or as LLVM code
blocks and macros based on GCC's nested
functions~\cite{DBLP:conf/oopsla/HarrisAIM11}.

CPS conversion for imperative languages is less common, probably because
it is harder to implement and prove correct.  CPS conversion has been
applied at least to C~\cite{weave},
C++~\cite{DBLP:conf/usenix/Mazieres01}, and
Javascript~\cite{DBLP:conf/usenix/MyersCCL07}. To the best of our
knowledge, CPC is the only public implementation for the C language, as
well as the only one using lambda-lifting to avoid the runtime overhead
of environments~\cite{hosc2012}.

The main downside of these program transformation techniques is that CPS
conversion changes function signatures, which makes it harder to mix
concurrent functions with external libraries expecting callbacks.
Unsurprisingly, similar issues arise when using events and threads
simultaneously; Adya et al.\ show how to use adaptors to connect both
styles~\cite{DBLP:conf/usenix/AdyaHTBD02}.

\paragraph{Static verification of real-world programs}

The constraint that only cooperative functions can call cooperative
functions is a very natural and common one in concurrent systems.  In
functional languages with a static type-checking, it is generally
enforced by the monadic structure or the type system
itself~\cite{DBLP:journals/jfp/Claessen99,DBLP:conf/ml/Vouillon08,DBLP:conf/icfp/RompfMO09}.
Interestingly enough, authors of similar systems for imperative
languages commonly acknowledge that static checking would be preferable,
but do not implement it~\cite{DBLP:conf/usenix/AdyaHTBD02,DBLP:conf/oopsla/HarrisAIM11}.
It seems that Kilim statically checks \verb+@pausable+ annotations,
although the authors do not mention it explicitly~\cite{DBLP:conf/ecoop/SrinivasanM08}.

\enlargethispage{-\baselineskip}
There is a long history of static analysis to enforce safety properties,
in particular for real-world programs written in languages lacking a
strong type
system~\cite{DBLP:conf/pldi/ConditHMNW03,DBLP:conf/sefm/CuoqKKPSY12}.
However, most of them require to add explicit annotations in ad-hoc
domain-specific languages.  A noteworthy exception is Dialyzer, a static
analyser reusing the annotation format already found in the
documentation of many Erlang programs~\cite{DBLP:conf/erlang/SagonasL08}.

\section{QEMU coroutines}
\label{sec:qemu-api}

As described in Section~\ref{sec:intro}, QEMU is an open source
machine emulator and virtualizer using a hybrid architecture that
combines event-driven programming with threads. The use of threads
mitigates two well-known limitations of event-driven architectures: an
event loop cannot take advantage of multiple cores because it only has a
single thread of execution; and long-running computations or blocking
system calls freeze every task in the event loop, not only the current
one.  Nevertheless, the core of QEMU is event-driven and most code
executes in that environment.

The main event loop is executed by a dedicated thread, called
\emph{iothread}.  When a file descriptor becomes ready, or when a timer
expires, it invokes a callback that responds to the event.  On the other
hand, guest code is executed by a number of \emph{vcpu} threads, one per
virtual (emulated) CPU. In addition, \emph{worker} threads are also used
to offload blocking operations outside of the main loop.

In 2011, in response to an increase of complexity in asynchronous code,
QEMU developers began to use coroutines to run concurrent tasks in the
iothread without splitting them into individual callback
functions~\cite{Wolf2011}.  A coroutine has its own stack and is
therefore able to preserve state across blocking operations, which
traditionally require callback functions and manual marshalling of
parameters.  Coroutines are now used heavily in the \emph{block layer},
the subsystem that provides access to disk image files and supports
background operations like live storage migration.  Coroutines allow
tasks requiring multiple disk updates to be expressed as sequential code
rather than breaking them (in event-driven style) into many functions
and explicitly passing on local variables.

\subsection{Coroutine API}

QEMU is written in C.  Since the C programming language does not include
support for coroutines, QEMU uses its own implementation that is based
on two annotations, two type definitions and five functions
(Figure~\ref{fig:qemu-api}).
\begin{figure}[htb]
  \begin{small}
\begin{verbatim}
#define coroutine_fn  /* implementation-dependent */
#define blocking_fn   /* implementation-dependent */
typedef struct Coroutine Coroutine;
typedef void coroutine_fn CoroutineEntry(void *);
Coroutine *qemu_coroutine_create(CoroutineEntry *);
void qemu_coroutine_enter (Coroutine *, void *);
void coroutine_fn qemu_coroutine_yield(void);
bool qemu_in_coroutine(void);
Coroutine * coroutine_fn qemu_coroutine_self(void);
\end{verbatim}
\end{small}
\caption{Coroutine interface in QEMU}
\label{fig:qemu-api}
\end{figure}

\enlargethispage{-\baselineskip}
Creating and starting a coroutine is very straightforward:
\begin{small}
\begin{verbatim}
coroutine = qemu_coroutine_create(my_coroutine);
qemu_coroutine_enter(coroutine, my_data);
\end{verbatim}
\end{small}
The function \verb+qemu_coroutine_create+ takes an entry function that
will be run inside a new coroutine, and returns a pointer to an opaque
structure \verb+Coroutine+, or \emph{coroutine handler}.  The entry
function must be of type \verb+CoroutineEntry+, i.e.\ taking an opaque
\verb+void*+ pointer and returning nothing.  The function
\verb+qemu_coroutine_enter+ transfers control to the coroutine.

The coroutine then executes until it returns, in which case it is
automatically freed, or yields:
\begin{small}
\begin{verbatim}
void coroutine_fn my_coroutine(void *opaque) {
  MyData *my_data = opaque;
  /* ... do some work ... */
  qemu_coroutine_yield();
  /* ... do some more work ... */
}
\end{verbatim}
\end{small}
Yielding is done either directly by calling \verb+qemu_coroutine_yield+, or
indirectly by calling a function that yields (itself directly or indirectly).
Yielding switches control back to the caller of \verb+qemu_coroutine_enter+.
This is typically used to switch back to the main thread's event loop after
issuing an asynchronous I/O request.  The request callback will then invoke
the function \verb+qemu_coroutine_enter+ once more to switch back to the
coroutine.

Finally, the QEMU coroutine interface provides a simple introspection
mechanism based on two functions: \verb+qemu_in_coroutine+ can be used
to check if the current function is executed in coroutine context, and
\verb+qemu_coroutine_self+ to get a pointer to the current coroutine if
this is the case.

\subsection{Coroutine and blocking annotations}
\label{sec:coro-annot}

Functions that are run inside a coroutine and may yield are called
\emph{coroutine functions}, and annotated with \verb+coroutine_fn+.
Note that any function that calls a coroutine function is prone to
yielding itself.  Therefore, a coroutine function may only be called by
another coroutine function; in other words, it is forbidden to call a
coroutine function from a non-coroutine, \emph{native} function.
Coroutine functions, on the other hand, are allowed to call native
functions.  Coroutine annotations are used twice in the coroutine API
(Figure~\ref{sec:qemu-api}):  coroutine entry points
(\verb+CoroutineEntry+) must be coroutine functions, since they are
executed in a coroutine, and \verb+qemu_coroutine_yield+ is of course
annotated as a coroutine function.

\emph{Blocking functions}, on the other hand, are native functions that
must not be called from a coroutine; they are annotated with
\verb+blocking_fn+.  We introduce this new annotation, which did not
exist in QEMU before this work, to identify native functions that
could block the main event loop for a long time and have a coroutine
equivalent that should be used instead.  Note that in principle, it
would be even safer to consider every native function as potentially
blocking, and annotate explicitly those that we wish to allow in
coroutine context; however, such a white-list mechanism would be
intractable in practice on a project of the size of QEMU, and we opted
for a black list of blocking functions instead.

One important limitation of QEMU before our work was that these global
constraints on the function call graph were not enforced in any way: as
shown in Figure~\ref{fig:qemu-api}, \verb+coroutine_fn+ and
\verb+blocking_fn+ are simply defined as empty macros by default, hence
discarded from the final source-code by the C preprocessor.  We discuss
in Sections~\ref{sec:cps-calling-convention} and~\ref{sec:corocheck} how
to give them a rigorous semantics, and how to check that annotated
functions are used correctly.

\subsection{Indirect coroutine calls}
\label{sec:indirect-calls}

QEMU uses indirect coroutine calls and function pointers to coroutine
functions intensively.  We have seen in Figure~\ref{fig:qemu-api} that
every call to \verb+qemu_coroutine_create+ involves a pointer to a
coroutine function, but this is far from the only place where they are
used.

As explained above, coroutines are mainly used to provide
non-blocking accesses to emulated disk images in the iothread.  To
support multiple disk image formats in an extensible way, the block
layer of QEMU defines a generic block driver interface
(Figure~\ref{fig:blockdriver}).  This interface consists in a set of
more than 40 callback functions that each driver needs to implement;
among them, 17 are coroutine functions.
\begin{figure}[htb]
  \begin{small}
\begin{verbatim}
struct BlockDriver {
  const char *format_name;
  int (*bdrv_probe_device)(const char *filename);
  int coroutine_fn (*bdrv_co_flush_to_os)
                          (BlockDriverState *bs);
  /* ... */
};
\end{verbatim}
\end{small}
\caption{Native and coroutine callbacks in block driver interface}
\label{fig:blockdriver}
\end{figure}

The block layer then uses this abstract interface to implement I/O
operations, performing many indirect calls to coroutine functions
provided by each driver. For instance, the coroutine function
\verb+brdv_co_flush+ calls the coroutine callback
\texttt{brdv\ub{}co\ub{}flush\ub{}to\ub{}os}.
\begin{small}
\begin{verbatim}
int coroutine_fn
bdrv_co_flush(BlockDriverState *bs)
{
  /* Write back cached data to the OS */
  if (bs->drv->bdrv_co_flush_to_os) {
   int ret = bs->drv->bdrv_co_flush_to_os(bs);
   if (ret < 0) {
     return ret;
   }
  }
  /* ... */
}
\end{verbatim}
\end{small}
To preserve the coroutine constraint on the call graph, it is essential
that function pointers be explicitly annotated.  Coroutine annotations
on function declarations and definitions alone are not enough to ensure
the correctness of coroutine calls.

\section{Static analysis of coroutine annotations}
\label{sec:corocheck}

Coroutine annotations are not only useful to document which functions
might block the event loop, and which ones can safely be used in a
non-blocking way.  In an event-driven program, as well as in a
concurrent system with cooperative threads or coroutines, the main loop
acts as a global lock, and it is common for programmers to rely on it to
synchronise access to shared resources or preserve global invariants.
Even in a stack-switching approach to coroutines, calling a coroutine
function outside of coroutine context can then lead to serious bugs.

Ironically, such a bug occurred in QEMU during the course of our
study. The code responsible for throttling disk I/O was causing a
segmentation fault because a function to reschedule coroutines,
\verb+qemu_co_queue_next+, was missing a coroutine annotation and was
called from native functions.  It remained broken for two months before
Canet identified the bug and fixed it~\cite{Canet2013}.
However, as we discovered later when checking statically the impacted
file, the fix itself still misses some coroutine annotations: getting
all of them correct without some form of automated verification is a
daunting task  (see Section~\ref{sec:coro-example} for more details).

The CPC translator enforces this rule statically, because it needs
correct annotations to drive its transformation.  However, it only
performs limited checking and is not convenient to analyse and infer
coroutine annotations on a large scale
(Section~\ref{sec:missing-annot}).  The fact that it interleaves the
analysis of coroutine annotations with source-to-source transformations
increases the opportunities for bugs, and makes it harder to add new
features.  In order to fix a large number of annotations in an efficient
and reliable way, we decided to write CoroCheck, a generic tool for the
static analysis and inference of coroutine annotations, designed to be
usable for CPC, QEMU, or any C other library with coroutine annotations
(Section~\ref{sec:coro-features}).

We illustrate the use of CoroCheck on a small example from QEMU in
Section~\ref{sec:coro-example}, and evaluate the number of annotations
that CoroCheck enabled us to fix in QEMU in Section~\ref{sec:coro-eval}.

\subsection{Missing coroutine annotations}
\label{sec:missing-annot}

In theory, it should have been enough to change a single line in the
header file \texttt{coroutine.h} to make the CPC translator recognize
and convert coroutine functions in QEMU:\footnote{See
Section~\ref{sec:cps-calling-convention} for an explanation of
\texttt{\_\_attribute\_\_((cps))}.}
\begin{small}
\begin{verbatim}
#define coroutine_fn __attribute__((cps))
\end{verbatim}
\end{small}
In practice, however, this early attempt failed because many coroutine
annotations in QEMU were missing, and CPC produced hundreds of errors
caused by inconsistent annotations.  As shown in
Table~\ref{tab:annot-stats} (Section~\ref{sec:coro-eval}, more than 70\,\%
of the coroutine annotations required to compile QEMU were missing.  In
hindsight, this should have come as no surprise.  As explained in
Section~\ref{sec:qemu-api}, coroutine annotations are used exclusively
for documentation purposes.  QEMU is a very large project, and not every
contributor understands how the coroutine mechanism works.  Even fewer
keep in mind the rules about coroutine annotations, and mistakes easily
go unnoticed since there is no automated check to detect them.

A naive approach to fix coroutine annotations is to blindly follow the
error messages reported by the CPC translator: adding missing
annotations where errors are reported, which will produce more errors at the
call points of these newly annotated functions, and iterating until one
reaches a fixed-point.   Apart from being extremely tedious, there are
two reasons why this naive approach does not work well, or even at all
in the case of QEMU: spurious annotations and hybrid functions.

\paragraph{Spurious annotations}

There is a risk of introducing too many annotations.  Each illegal call
to a coroutine function from a native function can be fixed either by
annotating the caller, or by wrapping the callee in a native function
allocating a dedicated coroutine for this call.  This is a design
choice that only the programmer can make, based on the concurrent
structure of the program: it would be unreasonable to systematically add
annotations all the way to the \texttt{main} function at the root of the
call graph because of a single annotation at one of the leaves.

Conversely, it is sometimes a deliberate choice to spuriously
annotate functions which do not call any other coroutine functions,
e.g.\ for documenting the intention and further plans to add cooperation
in some place of the code.  Hence spurious annotations should not be
removed automatically.

\paragraph{Hybrid functions}
Hybrid functions use \verb+qemu_in_coroutine+ to
check dynamically whether they are called in coroutine context, and
execute a different code path in each case.  Such functions do not work
with CPC, and need to be rewritten to split the coroutine and native
code paths.

The need for blocking functions (Section~\ref{sec:coro-annot}) arose
mainly when splitting hybrid functions into a native and a cooperative
version.  Annotating the native one as a blocking function would make
sure that it is not called by mistake from a coroutine, which would
block the whole event loop.  Both the splitting of hybrid functions and
the introduction of the \verb+blocking_fn+ annotation were discussed
with and approved by QEMU developers --- several of them considered
hybrid functions as a temporary work-around, used to ease the transition
when coroutines were first introduced in QEMU.

\subsection{CoroCheck}
\label{sec:coro-features}

CoroCheck is a generic tool for the static analysis and inference of
coroutine annotations, written in OCaml.  To make CoroCheck easily available and usable by
as many QEMU developers as possible, we extended CIL with a modular
plug-in system, and wrote CoroCheck as a CIL plug-in.  This improves the
previous cumbersome CIL architecture, which required users to recompile
their own version to add new features, and should make it easier for
every programmer to distribute analysis and program transformation tools
based on CIL in the future.

CoroCheck analyses one C file at a time. It assumes that coroutine functions and
function pointers are marked with an attribute,\footnote{The name of the
attribute is \texttt{coroutine\_fn} by default, and configurable with a
command-line option.} as detailed in
Section~\ref{sec:cps-calling-convention}, and that
the prototypes of functions implemented outside of the analysed file are
correctly annotated. This is a necessary assumption, since CoroCheck has no
means to determine whether extern functions are actually cooperative or not.

Assuming the correctness of external annotations implies that the
programmer is responsible for analysing the files in a topological order
based on their dependencies, or to iterate the analysis over the whole
project until reaching a fixed-point.  This proved not to be a problem
in practice.

For each file that it analyses, CoroCheck performs a coroutine annotation
inference, prints warnings for missing and spurious annotations, outputs an
annotated call graph to help the programmer analyse and fix those errors, and
checks that type casts and assignments respect coroutine annotations. We detail
each of these steps in the rest of this section.

\paragraph{Coroutine annotation inference}

CoroCheck builds a directed graph of the functions calls in the analysed file,
with a node for each function (either defined in the file, or simply declared),
and edges from caller to callee. It then uses the \texttt{Fixpoint} module of
the Ocamlgraph library~\cite{DBLP:conf/sfp/ConchonFS07} to propagate
coroutine annotations. As explained above, we start with the coroutine functions
implemented outside of the current file as a trusted root, and propagate the
annotations backwards to their callers until we reach a fixed-point.

There is in fact another class of functions that we add to the roots of the
algorithm: coroutine functions which have their address retained in a function
pointer.  This turned out to be necessary to avoid generating too many warnings
about spurious annotations: when a function is used to implement an interface
such as \texttt{BlockDriver} (Figure~\ref{fig:blockdriver}), it is essential to
obey the annotation constraints mandated by the interface. For further safety,
CoroCheck also verifies automatically that these function pointers are
used consistently.

CoroCheck then iterates over every function defined in the file and
prints warnings when the inferred coroutine annotation does not match
the original one.  It also warns if a blocking function is inferred as
cooperative, or called from a coroutine function (either inferred or
annotated in the original file).

\paragraph{Annotated call graph}

As explained in Section~\ref{sec:missing-annot}, blindly following CoroCheck
suggestions for annotating coroutine functions is not necessarily enough: one
might need to refactor an interface or split a hybrid
function for example.  Having a per-file graphical view of the function call graph proves
very helpful in understanding the reasons for the suggested fixes and analysing the
root causes of erroneous annotations.

\begin{figure*}[htb]
\begin{center}
\includegraphics[width=140mm]{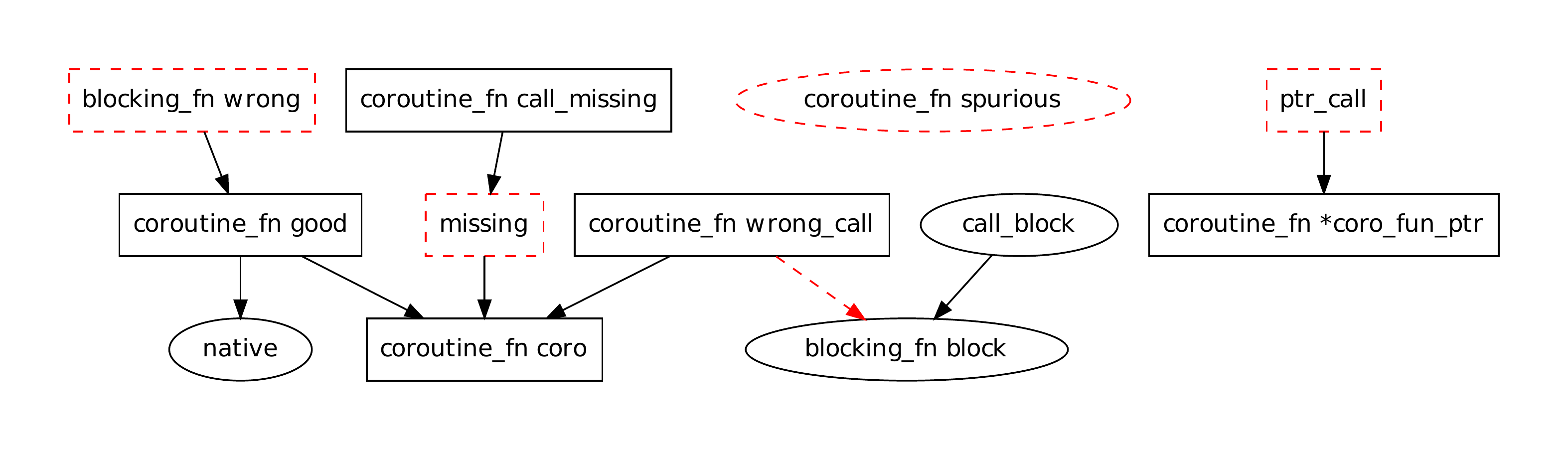}
\end{center}
\vspace{-5mm}
\caption{Function call graph annotated by CoroCheck}
\label{fig:corocheck}
\end{figure*}

CoroCheck uses Ocamlgraph facilities to output a file that can be
processed with the \texttt{dot} utility (from the GraphViz project) to
generate an image representing the annotated function call graph.
Consider for example the input program in Figure~\ref{fig:inference}; it
yields the output shown in Figure~\ref{fig:corocheck}.
\begin{figure}[htb]
\begin{small}
\begin{verbatim}
extern void coroutine_fn coro();
extern void blocking_fn block();

void native() { };
void coroutine_fn (*coro_fun_ptr)(void) = &coro;

void coroutine_fn spurious()     { }
void coroutine_fn good()         { coro(); native(); }
void              missing()      { coro(); }
void coroutine_fn call_missing() { missing(); }
void blocking_fn  wrong()        { good(); }
void              call_block()   { block(); }
void coroutine_fn wrong_call()   { coro(); block(); }
void              ptr_call()     { coro_fun_ptr(); }
\end{verbatim}
\end{small}
\caption{Input program for Figure~\ref{fig:corocheck}}
\label{fig:inference}
\end{figure}

Functions \emph{inferred} as coroutine functions are represented in
rectangles; native functions in ovals.  Annotations \emph{provided by
the programmer}, on the other hand, are printed within the boxes
(\texttt{coroutine\_fn} or \texttt{blocking\_fn} in
Figure~\ref{fig:corocheck}).  For indirect calls, since the name of the
called function is unknown, we print the expression used to perform the
call instead (eg.\ \texttt{coroutine *coro\_fun\_ptr}).

Mismatching nodes, corresponding to either spurious or missing
annotations, are ouput with dashed, red lines. For example, from
top-left to bottom-right in Figure~\ref{fig:corocheck}, the function
\texttt{wrong} should be annotated with \verb+coroutine_fn+
instead of \verb+blocking_fn+ because it calls the coroutine function
\verb+good+; the function \verb+spurious+ does not call any coroutine
function, so it does not need a \verb+coroutine_fn+ annotation; the
function \verb+ptr_call+ should be a coroutine functions because it
performs an indirect call to a coroutine function through the function
pointer \verb+coro_fun_ptr+; and the function \verb+missing+ should be
annotated as well since it calls the coroutine function \verb+coro+.
Note that the function \verb+call_missing+ is not flagged as spurious,
since it calls \verb+missing+ which should be a coroutine function (even
though it lacks the proper annotation).  Finally, dashed, red lines are also
used for forbidden edges from coroutine functions to blocking functions,
such as \verb+wrong_call+ calling \verb+block+.

Files from a real-world program can contain a huge number of functions, making
the call graph cluttered and challenging to decipher.  To focus on the relevant
information, CoroCheck removes every native function defined outside of
the current file from the graph. This strategy removes many leaves, in
particular all functions from the \texttt{libc} standard C library.  For
example, when applied to \texttt{block.c}, it removes more than half of
the nodes and edges (from 657~nodes and 781~edges down to 316~nodes and
302~edges).


\paragraph{Type cast and assignment verification}

As an additional safety check, unrelated to call graph analysis, CoroCheck
ensures that coroutine annotations are not lost when function pointers are
typecast or assigned to a variable.  This is relatively straightforward
thanks to CIL making every type cast explicit in its intermediary AST. 


\subsection{An example of coroutine-safety violation}
\label{sec:coro-example}

In this section, we show that annotations are significant even in the
original QEMU, and that a missing annotation can cause (and has in fact
caused) a serious and hard-to-find bug.

\begin{figure*}[htb]
\begin{center}
  \includegraphics[width=180mm]{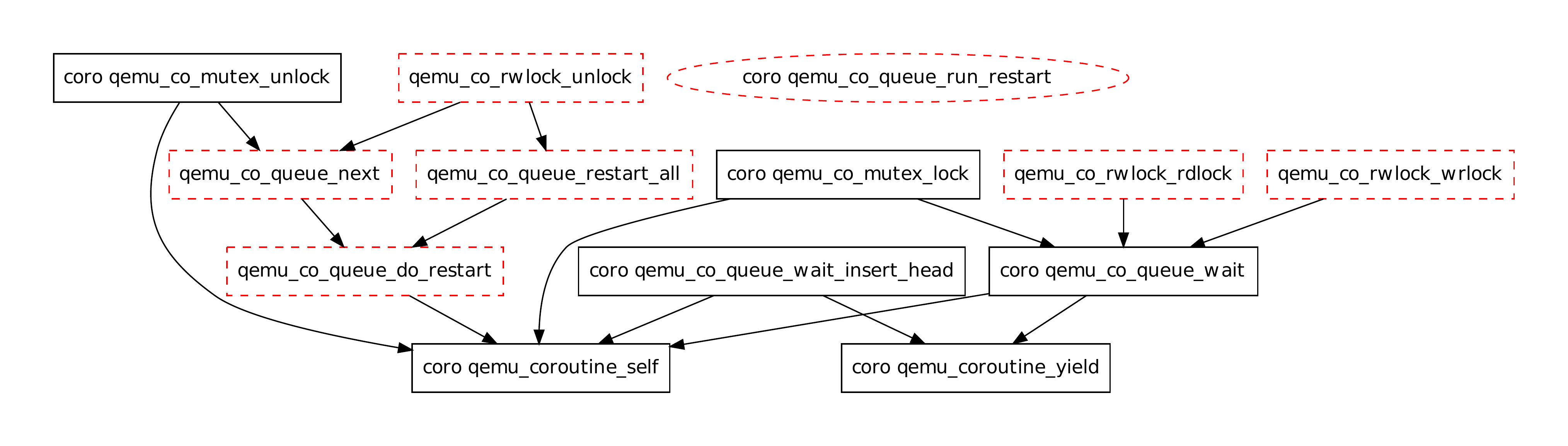}\\
  (a) Before Canet's patch\par
  \includegraphics[width=180mm]{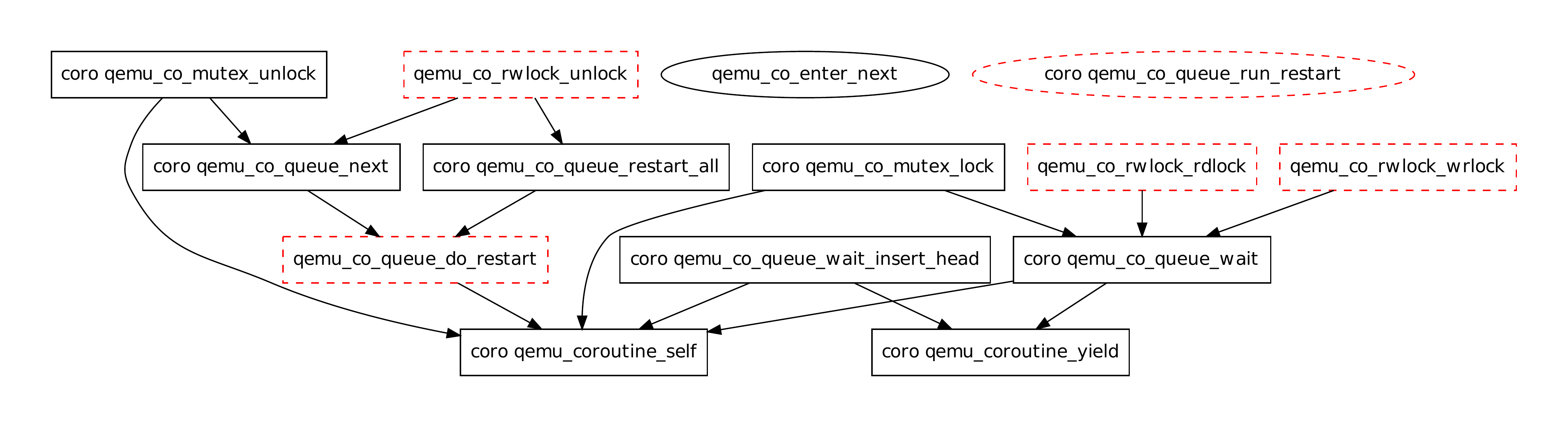}\\
  (b) After Canet's patch
  \par
  Note: the graph has been slightly simplified, and we use \texttt{coro}
  instead of \texttt{coroutine\_fn} to keep it readable.
\end{center}
\caption{Function call graph from \texttt{qemu-coroutine-lock.c},
annotated by CoroCheck}
\label{fig:lock}
\end{figure*}

Coroutine locks (\texttt{CoMutex}) are built on top a coroutine queues
(\texttt{CoQueue}).  The basic operations are to enqueue the current
coroutine, which transfers control to its caller
(\verb+qemu_co_queue_wait+), and to restart the next coroutine in a
queue (\verb+qemu_co_queue_next+). Locking operations are thin wrappers
around these functions which check if the \verb+CoMutex+ is already
locked before proceeding (\verb+qemu_co_mutex_lock+ and
\verb+qemu_co_mutex_unlock+). Figure~\ref{fig:lock}a gives an overview
of the file implementing coroutine queues and locks.

CoroCheck detects no less than six functions missing a coroutine
annotation, and one (\verb+qemu_co_queue_run_restart+) spuriously
annotated.  These missing annotations caused a serious bug: the code
responsible for throttling disk I/O trusted them, and called
\verb+qemu_co_next+ in native context. This ultimately led to calling
\verb+qemu_coroutine_self+ in native context, which returned an invalid
coroutine pointer and caused a segmentation fault in
\verb+qemu_co_queue_do_restart+.

Canet~\cite{Canet2013} fixed the segmentation fault two months later by
introducing \verb+qemu_co_enter_next+, a native-mode counterpart to
\verb+qemu_co_queue_next+.  However, as shown on Figure~\ref{fig:lock}b,
quite a few annotations are still missing or incorrectly placed after the
patch.  Without an automated verification tool such as CoroCheck, it is
probably only a matter of time before the wrong function is used again
by mistake in another part of QEMU.

\subsection{Evaluation}
\label{sec:coro-eval}

We have no doubt that it would have been much less likely to obtain
fully correct annotations in the short time-frame of our study without
using CoroCheck. Although this is but a subjective impression, the
figures shown in Table~\ref{tab:annot-stats} give a better idea of the
amount of work required.  Compared to the correctly annotated
\texttt{cpc} branch, the original \texttt{master} branch lacks more than
70\,\% of annotations.  We firmly believe that bug-free annotations
cannot be reached in a project of this scale without automated, static
verification tools.
\begin{table}[htb]
\caption{Coroutine annotations in QEMU}
\label{tab:annot-stats}
{\centering
\begin{threeparttable}
\begin{tabular}{llrr}
\toprule
QEMU branch&&\texttt{master}&\texttt{cpc}\\
\midrule                                                          
\texttt{coroutine\_fn} & in \texttt{.h} files &    31  &   100\\
                       & in \texttt{.c} files &   127  &   421\\
Hybrid functions       & (\texttt{qemu\_in\_coroutine}) &     6  &     1\\
\bottomrule
\end{tabular}
\begin{tablenotes}
\item
  A few files non-essential files have not been annotated yet, hence the
  remaining hybrid function.
\end{tablenotes}
\end{threeparttable}
}
\end{table}

\section{Continuation-Passing C}
\label{sec:cpc}

Continuation-Passing C (CPC)~\cite{hosc2012} is an extension of the C
language to write concurrent programs. The programmer writes synchronous
code in threaded style, using common synchronisation techniques such as
condition variables. This code is then automatically transformed into an
equivalent standard C program written in asynchronous, event-driven
style. Our previous experiments on small to medium-sized programs have
shown that the translated code is at least as efficient as stack-based
thread implementations, while providing a significantly smaller memory
footprint~\cite{Kerneis2012,places2011}.  However, CPC had never been
tested on programs as large as QEMU before this case study.

The CPC programmer uses a superset of the C language that we call the
\emph{CPC language}. It extends C with two keywords: \verb+cps+ to annotate
cooperative functions, and \verb+cpc_spawn+ to execute them in a new thread;
it implements half-a-dozen synchronisation primitives (\verb+cpc_yield+,
\verb+cpc_sleep+, etc.); and it provides a moderate amount of syntactic sugar
to write common concurrency idioms in a concise way.
As we have seen in Section~\ref{sec:intro}, a CPC program is compiled in
two steps:
\begin{enumerate}
  \item The \emph{CPC translator} first applies a series of
    source-to-source transformations to convert the original program
    into an equivalent one written in continuation-passing style.
    The resulting code is compiled by a C compiler, such as GCC.
  \item This compiled code is then linked with the \emph{CPC runtime},
    or, in the context of this paper, with a new coroutine backend for
    QEMU based on continuations (Section~\ref{sec:cpc-backend}).
\end{enumerate}

In the rest of this section, we give an overview of the compilation
technique used by the CPC translator on a small example. More details,
including proofs of correctness, are available in a previous
article~\cite{hosc2012}.

\subsection{The CPC compilation technique}

The CPC translator is structured in a series of proven source-to-source
transformations, which turn a threaded-style CPC program into an
equivalent event-driven C program.  \emph{Boxing} first encapsulates a
small number of variables in environments.  \emph{Splitting} then splits
the flow of control of each annotated function into a set of nested functions.
\emph{Lambda lifting} removes free local variables introduced by the
splitting step; it copies them from one nested function to the next,
yielding closed nested functions.  Finally, the program is in a form
simple enough to perform a one-pass partial \emph{CPS conversion}.  The
resulting continuations are used at runtime to schedule threads.

Consider the following function, which counts seconds down from an initial
value~\texttt{x} to zero.
\begin{small}
\begin{verbatim}
cps void countdown(int x) {
  while(x > 0) {
    printf("%d\n", x--);
    cpc_sleep(1);
  }
  printf("time is over!\n");
}
\end{verbatim}
\end{small}
This function is annotated with the \texttt{cps} keyword to indicate that it
yields to the CPC scheduler.  This is necessary because it calls the CPC
primitive \texttt{cpc\_sleep}, which also yields to the scheduler.  In
the rest of this article, we call such functions \emph{cps functions}.

We show below how splitting, lambda lifting and CPS conversion transform
the function \texttt{countdown}.  The boxing pass has no effect on this
example because it only applies to address-taken variables (the address
of which is retained by the ``address of'' operator \verb+&+).  It is
necessary in the general case to transform address-taken stack variables
into heap variables, because lambda-lifting and CPS-conversion passes
duplicate stack variables, which invalidates pointers storing their
address.

\subsection{Splitting}
\label{sec:splitting}

The next transformation performed by the CPC translator is \emph{splitting}.
Splitting has been first described by van Wijngaarden for Algol~60~\cite{wijngaarden}.
It translates control structures into mutually recursive
functions.

Splitting is done in two steps.  The first step consists in replacing every
control-flow structure, such as \texttt{for} and \texttt{while} loops, by its
equivalent in terms of \texttt{if} and \texttt{goto}.
\begin{small}
\begin{verbatim}
cps void countdown(int x) {
 loop:
  if(x <= 0) goto timeout;
  printf("%d\n", x--);
  cpc_sleep(1);
  goto loop;
 timeout:
  printf("time is over!\n");
}
\end{verbatim}
\end{small}
The second step uses the fact that \texttt{goto} are equivalent to tail calls~\cite{Ste76}.
It translates every labelled block into an nested function, and
every jump to that label into a tail call (followed by a \texttt{return}) to
that function.
\begin{small}
\begin{verbatim}
cps void countdown(int x) {
  cps void loop() {
    if(x <= 0) { timeout(); return; }        /* (1) */
    printf("%d\n", x--);
    cpc_sleep(1); loop(); return;            /* (2) */
  }
  cps void timeout() {
    printf("time is over!\n"); return;       /* (3) */
  }
  loop(); return;
}
\end{verbatim}
\end{small}

Splitting yields a program where each cps function is split in several mutually
recursive, atomic functions, very similar to event handlers.  Additionally, the
tail positions of these nested functions are always one of the following
three cases (numbered in the previous example):
\begin{enumerate}
  \item a tail call to another cps function (eg.\ \texttt{timeout}),
  \item a call to an external cps function (\texttt{cpc\_sleep})
    followed by a tail call to an nested cps function (\texttt{loop}),
  \item or a single \texttt{return} statement with no preceding call to
    a cps function.
\end{enumerate}
This restricted form, that we call \emph{CPS convertible} form, provides
strong enough guarantees to enable a correct and straightforward CPS
conversion at a later stage.

Another effect of splitting is that variables bound in the original
outer function appear free in the nested ones.  For instance, the
variable \texttt{x} is free in the function \texttt{loop} above.
Because C does not support nested functions, we need a pass of lambda
lifting to eliminate those free variables.

\subsection{Lambda lifting}

The CPC translator makes the data flow explicit with a lambda-lifting pass.
Lambda lifting, also called closure conversion, is a standard technique
introduced by Johnsson~\cite{DBLP:conf/fpca/Johnsson85} to remove free
variables.
It is also performed in two steps: parameter lifting and block floating.

Parameter lifting binds every free variable to the nested function where
it appears (for instance, \texttt{x} is bound to \texttt{loop} on line
\lineref{1} below).  The variable is also added as a parameter at every
call point of the function (lines \lineref{2} and \lineref{3}).
\begin{small}
\begin{verbatim}
cps void countdown(int x) {
  cps void loop(int x) {                     /* (1) */
    if(x <= 0) { timeout(); return; }
    printf("%d\n", x--);
    cpc_sleep(1); loop(x); return;           /* (2) */
  }
  cps void timeout() {
    printf("time is over!\n"); return;
  }
  loop(x); return;                           /* (3) */
}
\end{verbatim}
\end{small}
Block floating is then a trivial extraction of closed, nested functions
at top-level.

Note that because C is a call-by-value language, lifted parameters are
duplicated rather than shared. Therefore, this transformation is not
correct in general: mutating a copied parameter would leave the original
one intact, which could in principle be observed and yield different
results.  It is however sound in the case of CPC because lifted
functions are always called in tail position
(Section~\ref{sec:splitting}): they never return, which guarantees that
at most one copy of each parameter is reachable at any given
time~\cite[Section~6]{hosc2012}.


\subsection{CPS conversion}

Finally, the control flow is made explicit with a CPS
conversion~\cite{DBLP:journals/tcs/Plotkin75,strachey}.  The
continuations store callbacks and their parameters in a regular
stack-like structure \texttt{cont} with two primitive operations:
\texttt{push} to add a function on the continuation, and \texttt{invoke}
to call the first function of the continuation.
\begin{small}
\begin{verbatim}
cps void loop(int x, cont *k) {
  if(x <= 0) { timeout(k); return; }
  printf("%d\n", x--);
  cpc_sleep(1, push(loop, x, k)); return;
}
cps void timeout(cont *k) {
  printf("time is over!\n");
  invoke(k); return;
}
cps void countdown(int x, cont *k) {
  loop(x, k); return;
}
\end{verbatim}
\end{small}
Just like lambda lifting, CPS conversion is not correct in
general in an imperative call-by-value language, because variables are
duplicated to be stored on the continuation.  It is however correct in
the case of CPC, for reasons similar to the correctness of lambda
lifting~\cite[Chapter~5]{Kerneis2012}.
\section{CPS-converting QEMU}
\label{sec:cpc-qemu}

Our continuation-based backend for QEMU is made of two parts, similar to
the two-step approach used to compile CPC programs
(Section~\ref{sec:cpc}).  We first need to perform a CPS conversion of
every annotated coroutine function, and then use the continuations
introduced by the CPC translator to implement QEMU's coroutine API
(Figure~\ref{fig:qemu-api}).

In this section, we study the challenges associated with these two
steps: parsing correctly such a large-scale base of C code
(Section~\ref{sec:cil-bugs}), and defining CPS conversion as a C calling
convention to support indirect calls to CPS-converted functions
(Section~\ref{sec:cps-calling-convention}); then implementing the
runtime \texttt{coroutine-cpc} backend (Section~\ref{sec:cpc-backend}).

\subsection{Compiling QEMU with CIL}
\label{sec:cil-bugs}

CPC is built on top of the C Intermediate Language (CIL)
framework~\cite{DBLP:conf/cc/NeculaMRW02}.  CIL is a front-end for the C
programming language that facilitates program analysis and
transformation. It parses and typechecks an input program, and translate
it into an equivalent, simplified subset of C. In CIL, for example,
expressions have no side-effect and all looping constructs are
normalised to a \verb+while(1)+ loop with \verb+break+ statements.  The
programmer then manipulates this simplified AST, which greatly reduces
the number of cases that must be considered. CIL finally outputs the
resulting AST in a C file which is handed over to the actual C compiler.
A Perl script acts as a drop-in replacement for GCC, automating these
stages and making it easy to compile existing projects through CIL.

Before attempting to translate coroutine functions with CPC, we needed
to make sure that the CIL front-end was able to correctly parse and
translate QEMU source code.  Our very first step was therefore to try
and compile QEMU using CIL alone.  Although CIL is reasonably complete
and well-tested, QEMU is a large project with over 750\,000 lines of C
code, and it uses a wide range of C features, including a number of
non-standard GCC extensions.

While attempting to compile QEMU with CIL, we discovered and fixed at
least two major bugs, and several lacking features.  All those fixes and
features are now included in CIL and will be available in the next
release.

\paragraph{Major bugs in CIL}

The most disconcerting CIL bug that we encountered did not prevent QEMU
from compiling. Instead, it caused a crash of the guest operating system
when we tried to start a Linux~3.2 kernel with hardware virtualization
disabled. Older kernels seemed to work completely fine, as well as newer
kernels when virtualization was enabled. It turned out to be an
erroneous implementation in CIL of C rules for arithmetic conversion. This
fundamental bug had gone unnoticed for so long because it did not produce
observable effects except in some corner case on \texttt{long long}
integers, triggered in particular by QEMU's emulation of Intel's SSE
extensions.

We also found a bug in initialization of arrays: CIL discarded trailing
empty initializers. For instance, \texttt{\{\{.x = 3\}, \{.x = 5\}, \{\}
\}} was interpreted as an array of size 2 instead of 3; the last
initializer is equivalent to \texttt{\{.x = 0\}} in that case. This idiom is
used extensively in QEMU to mark the end of arbitrarily sized arrays of
complex structures without setting explicitly each field of the last
element to \texttt{0}.

A number of more minor bugs were also fixed along the way. The Perl
wrapper script in particular needed to be updated because it failed to
pass through some GCC options.

\paragraph{New C features and GCC extensions}

We added support for flexible array members, a C99 feature allowing an
array of unspecified size to be the last member of a C structure. This
is used in several places in QEMU, to allocate an array of data and some
meta-data in a compact, cache-efficient way.

We also improved support of GCC extensions. We added many new GCC
builtins, including concurrency operations for the C11 memory model.
We implemented first-class support for case-ranges, allowing to write
\texttt{switch} statements with conditionals of the form \texttt{case
0x0000 ...  0xc000}; CIL already accepted that extension, but translated
it into an exhaustive list of all cases, which did not scale to very
large ranges.

As an aside, some GCC extensions are not available on every platform,
and QEMU provides fall-back mechanisms for those cases.  This has
allowed us to disable two extensions that would have required an
unreasonable amount of work to be added to CIL: 128-bit integers
(\texttt{int128\_t}) and SSE vector instructions.

\subsection{Indirect coroutine calls}
\label{sec:cps-calling-convention}

The next stumbling block on the path to CPS conversion was indirect
coroutine calls.  In the original CPC implementation, the \texttt{cps}
annotation is a new keyword defined as a \emph{function specifier}. This
means that it applies to the function being annotated, rather than to
its type.  This unfortunate decision, taken at an early stage in CPC
design, forbade \texttt{cps} annotations at the type level, and made it
impossible to apply a CPS conversion to QEMU source code, because of the
pervasive use of coroutine function pointers
(Section~\ref{sec:indirect-calls}).  In this section, we argue that
coroutine annotations denote a \emph{calling convention}~\cite{ritchie},
and we detail how to implement this approach in CPC.

Native functions need to agree on a calling convention, often defined in
the Application Binary Interface (ABI) of the architecture they are
compiled for, for instance to decide how to pass function parameters (on
the stack or via register), or whether stack frames should be cleaned by
the caller or the callee.  Even given a particular language and
architecture, several calling conventions might coexist.  For instance,
on Intel~386, the \emph{cdecl} convention passes every argument on the
stack, whereas the \emph{fastcall} convention passes the first two
arguments in registers \texttt{ECX} and \texttt{EDX}.

There is no standard way in C to specify the calling convention
associated to a function or a function pointer.  Fortunately enough, to
ease interoperability, most compilers provide a means of specifying the
calling convention explicitly on a per function.  The \emph{de facto}
standard is to use attributes, a generic mechanism which allows to
annotate types, function declarations and even expressions.

Figure~\ref{fig:call-conv} shows how function attributes for calling
conventions are used in practice.  The attribute can be applied to a
function prototype, like \texttt{f} \lineref{1}, or to the type of a
function pointer like \texttt{p} \lineref{2}.  Tracking the calling
convention within the type of each pointer then allows to perform
indirect calls with the correct convention \lineref{4}.  The assignment
of a value to the pointer \lineref{3} and the implementation of a
prototype \lineref{5} are checked by the compiler, which will issue a
warning or an error if an incompatible calling convention is used; in
our example, line \lineref{3} is correct but \lineref{5} is forbidden.

\begin{figure}[htb]
\begin{small}
\begin{verbatim}
int __attribute__((fastcall)) f(int a, int b);    /* 1 */
int __attribute__((fastcall)) (*p)(int a, int b); /* 2 */
p = &f;                                           /* 3 */
int z = (*p)(1,2);                                /* 4 */
int f(int a, int b) { return a + b; }             /* 5 */
\end{verbatim}
\end{small}
\caption{Calling convention attributes}
\label{fig:call-conv}
\end{figure}

Those are exactly the properties that we need to keep track of coroutine
annotations within types, and to be able to perform indirect
CPS-converted calls.  As a matter of fact, CPC actually defines a calling
convention for cps functions, specifying how to store parameters on the
continuation and how to pass return values to the next function.  This
calling convention is precisely the set of assumptions used by the CPC
runtime to interface primitive functions with the code produced by the
CPC translator.

Implementing coroutine annotations with function attributes rather than
the \texttt{cps} keyword simplified the lexing and parsing stages, and
made the internal data structures and transformations used by the CPC
translator less ad-hoc.  Backward-compatibility merely required to add a
single line to the runtime header:
\begin{small}
\begin{verbatim}
#define cps __attribute__((cps))
\end{verbatim}
\end{small}

We encountered two technical difficulties.  First, we needed to be very
careful to be compatible with the way \texttt{gcc} and other compilers
implement calling-convention function attributes.  The rules for
positioning are very liberal, and not formally defined in the
documentation of compilers: in general, attributes apply to the nearest
component of the type, but in the case of calling convention attributes,
they can be placed almost anywhere within the return type and still
apply to the function type as a whole.  Then, calling-convention
attributes also need to be treated differently than other attributes
when merging the attributes of a prototype with those of the
implementation: because they create a constraint on the caller, it is
essential to ensure that the calling convention is the same in all cases
to ensure a consistent usage between the header and the implementation
(Figure~\ref{fig:call-conv}, line~\texttt{5}).

\subsection{The coroutine-cpc backend}
\label{sec:cpc-backend}

The last step of the QEMU/CPC project is \verb+coroutine-cpc+, a new
implementation of coroutines for QEMU based on the continuations
introduced by the CPC translator.

One important requirement was to write a simple, self-contained
backend: having as few changes to QEMU as possible, except for the
missing coroutine annotations that are dealt with separately, means that
it is easier to track the evolution of QEMU, and to potentially merge
our backend at some point in the future.  The implementation of
\verb+coroutine-cpc+ is short, with less than 200~lines of portable,
carefully-optimised code.  It also limits the amount of new code to the
minimum, by re-using code from the original CPC runtime implementation
for each low-level management task: allocating, deallocating, resizing
continuations, and passing a return value to the next
continuation~\cite[Section~3.2.4]{Kerneis2012}.

The QEMU coroutine API and the CPC runtime are very close, offering the
same kind of primitives.  One noticeable difference is that QEMU exposes
coroutines, providing an explicit handle for the programmer to schedule
them, whereas CPC exposes threads with an implicit scheduler and no
thread handle.  The interface of the former is therefore lower-level but
slightly more expressive than that of the latter.  As a result, fewer
native cps functions need to be implemented for the QEMU API than for
the CPC runtime: functions such as \verb+cpc_sleep+ or
\verb+cpc_io_wait+ need to hook into the CPC scheduler, whereas they are
built independently from the backend in the case of QEMU, on top the
basic coroutine API.  For the \texttt{coroutine-cpc} backend, we only need to
implement four QEMU-specific functions: \verb+qemu_coroutine_enter+,
\verb+qemu_coroutine_yield+, \verb+qemu_coroutine_self+ and
\texttt{qemu\ub{}in\ub{}coroutine}.

\paragraph{Entering and yielding coroutines}

The implementation of the function \verb+qemu_coroutine_enter+
initialises the continuation if this is the first time it is entered,
and starts a trampoline loop to run it (see Figure~\ref{fig:trampoline}).
The function pointer on top of the continuation is extracted
\lineref{2}, and called with the rest of the continuation as a unique
parameter \lineref{4}; it returns a new continuation, and the process is
repeated until reaching the empty continuation \lineref{1} or a call to
\verb+qemu_coroutine_yield+ \lineref{3}.
\begin{figure}[htb]
\begin{small}
\begin{verbatim}
while(1) {
  /* (1) If continuation is empty, return */
  if (k->length == 0) return k;
  /* (2) Otherwise, extract function pointer */
  k->length -= sizeof(cpc_function *);
  f = *(cpc_function **)(k->c + k->length);
  /* (3) Intercept yield if necessary */
  if (f == qemu_coroutine_yield) return k;
  /* (4) Otherwise, run the extracted function */
  k = (*f)(k);
}
\end{verbatim}
\end{small}
\caption{Trampoline loop for continuations}
\label{fig:trampoline}
\end{figure}

The usual way to implement yield in continuation-passing style is to
write the CPS-form directly by hand, returning a null pointer to
indicate that there is no continuation left to execute because the
coroutine has yielded (and adapting the trampoline loop accordingly).
This approach is not convenient in the case of QEMU because
\verb+qemu_coroutine_yield+ is defined in a supposedly
backend-independent way, which in fact relies implicitly on a
stack-switching implementation.  To work-around this limitation without
modifying files outside of \verb+coroutine-cpc+, we intercept the call
to \verb+qemu_coroutine_yield+ in the trampoline loop, by comparing
function pointers (Figure~\ref{fig:trampoline}).

\paragraph{Dynamic bookkeeping}

To implement the introspection functions \verb+qemu_coroutine_self+ and
\verb+qemu_in_coroutine+, all existing QEMU backends use a thread-local
variable to keep track of the current coroutine. This variable is
updated on each coroutine switch by \verb+qemu_coroutine_enter+.  As it
turns out, this dynamic bookkeeping is no longer necessary for
\verb+coroutine-cpc+.

It is enough to annotate these functions with the special attribute
\verb+cpc_need_cont+.  Then, the CPC translator passes them the current
continuation directly when they are called from coroutine context, and a
null pointer otherwise.  In fact, once hybrid functions have been
eliminated, \verb+qemu_in_coroutine+ becomes completely redundant with
coroutine annotations, its use being subsumed by the static verification
of CoroCheck.

In addition to making the code simpler and safer, we measured a speedup
of up to 10\,\% in micro-benchmarks when we removed dynamic bookkeeping
in \verb+coroutine-cpc+.

\section{Experimental results}
\label{sec:eval}

In this section, we compare the performance of coroutine backends on
micro-benchmarks (Section~\ref{sec:microbench}) and evaluate the
performance impact of CPS conversion on a typical QEMU work-load
(Section~\ref{sec:macrobench}).

\subsection{Micro-benchmarks}
\label{sec:microbench}

To evaluate the efficiency of basic coroutine operations, we use three
micro-benchmarks from the test suite of QEMU.\footnote{Our benchmarks
scripts are available online: \url{http://github.com/kerneis/cpc-qemu-bench/}.}
  \emph{Lifecycle} repeatedly creates an empty coroutine, which is entered then
    destroyed immediately.
  \emph{Nesting} repeatedly creates 1\,000 nested coroutines, each of them
    incrementing a shared counter, then creating and entering the next
    coroutine.
  \emph{Yield} repeatedly enters a single coroutine which decrements a counter,
    then yields immediately.

We test the continuation-based backend \texttt{cpc}, the stack-switching
backends \texttt{ucontext} and \texttt{sigaltstack}, and the thread-based
backend \texttt{gthread}.  All benchmarks are compiled directly with gcc,
except the \texttt{cpc} backend which uses the CPC translator; we have verified
that the results are unchanged when compiling the other backends with the CIL
front-end.  The results are shown in Table~\ref{tab:microbench}.

Because allocating coroutines is a costly process for most backends, QEMU uses
a pool of 64~coroutines: instead of freeing coroutines that have completed, they
are kept in a linked-list, and re-used when a new coroutine is needed.  To
measure the impact of allocation, we perform each benchmark with and without the
coroutine pool (except for \texttt{gthread}, which does not support the pool at
all).

\begin{table}[htb]
\caption{Speed of basic coroutine operations for various backends}
\label{tab:microbench}
{\centering
\begin{threeparttable}
\begin{tabular}{@{}lrrrrrr@{}}
\toprule
& \multicolumn{2}{c}{Lifecycle} & \multicolumn{2}{c}{Nesting} & \multicolumn{2}{c}{Yield} \\
Pool  & \multicolumn{1}{c}{no}    &     \multicolumn{1}{c}{yes} &
\multicolumn{1}{c}{no}    &     \multicolumn{1}{c}{yes} &  \multicolumn{1}{c}{no}
&     \multicolumn{1}{c}{yes} 
\tabularnewline                                                   
\midrule                                                          
            cpc &        75 &        54 &          94 &     125  &       19  &       19  \\ 
       ucontext &       464 &       108 &      3\,899 &     682  &       83  &       84  \\
    sigaltstack &    1\,796 &       108 &      5\,843 &  1\,988  &       85  &       87  \\
        gthread &   10\,802 &       --- & 2\,826\,905 &     ---  &   5\,703  &      ---  \\
\bottomrule
\end{tabular}
\begin{tablenotes}
\item All speeds are in nanoseconds, averaged over 10 runs (except
  \emph{gthread-nesting}, over 5 runs) of millions of iterations, on an 8-core
  Intel Xeon E5-1620 at 3.6~GHz.
  For \emph{nesting}, the time is per nested coroutine, hence
  comparable to \emph{lifecycle} directly.
\end{tablenotes}
\end{threeparttable}
}
\end{table}

The \texttt{cpc} backend is consistently faster than every other backend.
Allocation is faster because continuations are resized dynamically when needed,
whereas other backends need to allocate a large chunk of memory at once. As a
result, the coroutine pool is extremely effective for other backends, but turns
out to slow down \texttt{cpc} in the \emph{nesting} benchmark: when the number
of coroutines is one order of magnitude larger than the size of the pool, the
cost of managing the pool becomes higher than its benefits. As expected, the
pool has almost no impact for \emph{yield}, since only one coroutine is used in
this test.

We are not sure exactly why the gap is so large between the \emph{lifecycle} and
\emph{nesting} benchmarks for most backends: both tests perform exactly the
same task, except that the latter uses more coroutines simultaneously and
maintains a shared counter between coroutines.  The memory pressure is certainly
higher, but it is not clear why it slows down the creation of each
coroutine so much, especially when the pool is disabled anyway.  We believe that
cache effects are involved here, which would explain why \texttt{cpc} performs
much better with its small memory footprint.

The result of the \emph{yield} benchmark is not surprising. Coroutine switching
is a mere function return in the case of \texttt{cpc}, hence faster than the
signals, stack-switching mechanisms or locks used by the other backends.

\subsection{Macro-benchmarks}
\label{sec:macrobench}

The main downside of CPS conversion is that it adds an overhead to each
call to a CPS-converted function.  It is hard to predict the global
overhead on a large program because the splitting pass introduces a
number of calls to CPS-converted functions which varies with the
complexity of the control-flow and the position of cooperation points.
While the CPC translator tries to limit the number of inserted calls by
performing incremental transformations, we need to perform
macro-benchmarks to evaluate the overall effectiveness of our approach.

Since coroutines are mainly used in the block layer implementation, we
need to generate a lot of disk I/O from the virtual machine.  We use a
virtualized guest OS with Debian ``squeeze'', ran on a Debian
``squeeze'' host on an 8-core Intel Xeon 3.6~GHz, and \texttt{fio} to
generate various intensive disk accesses patterns, with up to
500~simultaneous readers or writers.  The guest is installed in a disk
image using QEMU's \texttt{qcow2} disk format.  We take particular
care to disable disk caches in both the guest OS and QEMU's disk layer,
to make sure each access in the guest translates to an actual read or
write of the image file on the host; disk cache is kept enabled on the
host.  We compare the mean and median access time for each access
pattern and coroutine backend.

Unfortunately, none of the coroutine backend performs significantly
faster or slower than the others in these macro-benchmarks.  To get a
finer-grained understanding of the results, we profile each QEMU
instance with the Linux \texttt{perf} utility, which uses hardware
performance counters provided by the CPU.  As it turns out, coroutines
are not at all on the critical path when emulating or virtualizing a
whole system: slightly less than 1\,\% of the execution time is spent in
coroutine-related functions, and the differences between backends end
up being smaller than the variability of disk-access times.

However, \texttt{perf} allows for a fine-grained analysis of the time
spent in each function, with some measurement uncertainty due to its
event-sampling approach.  Therefore, it is possible to isolate the most
time-consuming coroutine function in \texttt{perf} results (for instance
\texttt{qcow2\_co\_writev} for a write benchmark), and compare its
ranking for the various coroutine backends.

This is a tedious process that we did not manage to automate: since
the splitting pass of the CPC translator creates new coroutine
functions, one needs to sum those to recover the global time spent in
the original function.  In the few runs that we analysed, the coroutine
functions split by the CPC translator did not take significantly more
time to execute.  A likely explanation is that coroutine functions frequently yield
in practice; the time wasted in the trampoline loop is negligible
compared to the overhead of entering and switching coroutines for other
backends.

Despite our efforts, these macro-benchmarks are frustratingly
inconclusive.  Profiling at least seems to indicate that CPS conversion
does not generate a significant overhead in the case of QEMU.  This is
in line with the conclusions of our previous experiments on smaller
programs~\cite{places2011}.

\section{Conclusions}
\label{sec:conclusion}

We have applied a conversion to continuation-passing style to QEMU, a large
open-source project written in C with heavy use of function pointers.  Then, we
have used these continuations to implement an alternative coroutine backend,
both more portable and much faster than the existing ones. We have also
developed CoroCheck, a tool for the static analysis of coroutine annotations,
and used it to correct several hundreds of missing annotations in QEMU.

Our work demonstrates that static analysis of coroutines helps
ruling out some actual bugs. It also shows that CPC is flexible and
mature enough to scale efficiently to very large programs, and to a
model (coroutines) that it was not initially designed to handle.

Beyond the scientific results, this is the story of a successful collaboration
between an open-source project and academic research.  Over the course of a few
months, we have started a fruitful relationship, improving the state of
coroutines in QEMU, developing a tool that they can use as part of their test
suite, and fixing many bugs in CIL and CPC.  We were certainly not the first,
but we hope that many others will come after us, and discover the mutual
benefits of scaling-up their techniques to real-world projects.

\acks

The authors wish to thank Alan Mycroft and Gabriel Scherer for their thoughtful
suggestions on the structure of this article, as well as Matthieu Boutier, Marc
Lasson, Raphaël Proust, and Peter Sewell for their comments.
They are also beholden to the anonymous reviewers for their extremely
detailed and valuable suggestions.
This work has been supported by Google (Google Summer of Code 2013) and
EPSRC (grants EP/H005633 and EP/K008528).


\bibliographystyle{abbrvnat}


\end{document}